\documentclass[natbib,epj]{svjour}

\usepackage{amsmath}
\usepackage{amssymb}
\usepackage{amsfonts}
\usepackage{graphicx}
\usepackage{dcolumn}
\usepackage{bm}
\usepackage{color}
\usepackage{multirow,longtable}
\usepackage{times}

\begin{document}

\title{Multiscaling behavior in the volatility return intervals of Chinese indices}%

\author{Fei Ren\inst{1,2,}{\color{blue}{\thanks{e-mail:
fren@ecust.edu.cn}}} \and Wei-Xing
Zhou\inst{1,2,3,4,}{\color{blue}{\thanks{e-mail:
wxzhou@ecust.edu.cn}}}}

\institute{School of Business, East China University of Science and
Technology, Shanghai 200237, China \and Research Center for
Econophysics, East China University of Science and Technology,
Shanghai 200237, China \and School of Science, East China University
of Science and Technology, Shanghai 200237, China \and Research
Center of Systems Engineering, East China University of Science and
Technology, Shanghai 200237, China}

\titlerunning{Multiscaling behavior in volatility return intervals of Chinese indices}


\date{Received \today  / Revised version: }

\abstract{We investigate the probability distribution of the return
intervals $\tau$ between successive 1-min volatilities of two
Chinese indices exceeding a certain threshold $q$. The
Kolmogorov-Smirnov (KS) tests show that the two indices exhibit
multiscaling behavior in the distribution of $\tau$, which follows a
stretched exponential form $f_q(\tau/\langle \tau \rangle)\sim e^{-
a(\tau/ \langle \tau \rangle)^{\gamma}}$ with different correlation
exponent $\gamma$ for different threshold $q$, where $\langle \tau
\rangle$ is the mean return interval corresponding to a certain
value of $q$. An extended self-similarity analysis of the moments
provides further evidence of multiscaling in the return intervals.
}

\maketitle

\section{Introduction}
\label{intro}

The analysis of the waiting time between two successive events is
helpful to understand the dynamics of stock markets, which has drawn
much attention. A variety of waiting time variables have been raised
by different definitions of {\em{event}} to characterize the stock
markets from different view angles, such as the persistence
probability
\cite{Zheng-2002-MPLB,Ren-Zheng-2003-PLA,Ren-Zheng-Lin-Wen-Trimper-2005-PA},
the exit time
\cite{Simonsen-Jensen-Johansen-2002-EPJB,Jensen-Johansen-Simonsen-2003-IJMPC,Jensen-Johansen-Simonsen-2003-PA,Zhou-Yuan-2005-PA,ZaluskaKotur-Karpio-Orlowski-2006-APPB,Karpio-ZaluskaKotur-Orlowski-2007-PA},
and the intertrade duration
\cite{Scalas-Gorenflo-Luckock-Mainardi-Mantelli-Raberto-2004-QF,Ivanov-Yuen-Podobnik-Lee-2004-PRE,Sazuka-2007-PA,Jiang-Chen-Zhou-2008-PA}.
Recently the return intervals between successive extreme events
exceeding a certain threshold $q$ have been investigated for
numerous complex systems, including rainfalls, floods, temperatures
and earthquakes
\cite{Schmitt-Nicolis-2002-Fractals,Bunde-Eichner-Havlin-Kantelhardt-2003-PA,Bunde-Eichner-Havlin-Kantelhardt-2004-PA,Bunde-Eichner-Kantelhardt-Havlin-2005-PRL,Saichev-Sornette-2006-PRL}.
Similar analysis was subsequently carried out concerning the
volatility return intervals, which are defined as the waiting times
between successive volatilities exceeding a certain threshold in
stock markets.

Yamasaki {\em{et al.}} and Wang {\em{et al.}} used the daily data
and intraday data of US stocks to study the properties of the
volatility return intervals
\cite{Yamasaki-Muchnik-Havlin-Bunde-Stanley-2005-PNAS,Wang-Yamasaki-Havlin-Stanley-2006-PRE,Wang-Weber-Yamasaki-Havlin-Stanley-2007-EPJB,VodenskaChitkushev-Wang-Weber-Yamasaki-Havlin-Stanley-2008-EPJB}.
They found that the distribution of return intervals $\tau$ between
successive volatilities greater than a certain threshold $q$ showed
scaling behavior. This scaling behavior is expected to be of great
importance for the risk assessment of large price fluctuations.
Similar scaling behavior was observed in the return intervals of
daily and 1-min volatilities of thousands Japanese stocks
\cite{Jung-Wang-Havlin-Kaizoji-Moon-Stanley-2008-EPJB}. Qiu, Guo and
Chen analyzed the high-frequency intraday data of four liquid stocks
traded in the emerging Chinese market, and found that the return
interval distributions of the Chinese stocks investigated also
followed a scaling behavior \cite{Qiu-Guo-Chen-2008-PA}.

In contrast, Lee {\em{et al.}} investigated the return intervals of
1-min volatility data of the Korean KOSPI index
\cite{Lee-Lee-Rikvold-2006-JKPS} and no scaling was observed. Wang
{\em{et al.}} used the Trade \& Quate Database of the 500
constituent stocks composing the S\&P 500 Index and found a
multiscaling behavior in the volatility return intervals
\cite{Wang-Yamasaki-Havlin-Stanley-2008-PRE}. A systematic deviation
from scaling was observed in the cumulative distribution of return
intervals, which implies that its probability distribution also
deviates from scaling. Moreover, the $m$-th moment of the scaled
return intervals showed a certain trend with the mean interval,
which supports the finding that the return intervals exhibit
multiscaling behavior. This finding was reinforced by further
analysis of 1137 US common stocks
\cite{Wang-Yamasaki-Havlin-Stanley-2008-XXX}. Ren, Guo and Zhou used
a nice high-frequency database \cite{Jiang-Guo-Zhou-2007-EPJB} to
study the interval returns of 30 most liquid stocks in Chinese stock
market \cite{Ren-Guo-Zhou-2009-PA}. The Kolmogorov-Smirnov (KS) test
was adopted to examine the possible collapse of the interval
distributions for different threshold values. Only 12 individual
stocks passed the KS test and showed a scaling behavior, while the
remaining 18 stocks exhibited multiscaling behavior.

In this paper, we study the distribution of the volatility return
intervals of two Chinese stock indices, i.e., Shanghai Stock
Exchange Composite Index (SSEC) and Shenzhen Stock Exchange
Composite Index (SZCI). According to the KS test and the extended
self-similarity (ESS) analysis of the moments, we find that the
return intervals of the two indices exhibit multiscaling behavior,
consistent with the multiscaling behavior of some individual stocks
which partially compose the indices. The paper is organized as
follows. In Section \ref{S1:Data}, we explain the database analyzed
and how the volatility return intervals are calculated. Section
\ref{S2:KS} examines the scaling behavior and the curve fitting of
the return interval distributions using the KS tests. In Section
\ref{S5:Moment}, we further study the multiscaling behavior by
analyzing the moments of the scaled return intervals. Section
\ref{S8:concl} concludes.

\section{Preprocessing the data sets}
\label{S1:Data}

Our analysis is based on the high-frequency intraday data of two
Chinese indices, the Shanghai Stock Exchange Composite Index (SSEC)
and the Shenzhen Stock Exchange Composite Index (SZCI). Each
composite index is constructed based on all the stocks listed on the
corresponding exchange. The indices are recorded every six to eight
seconds from January 2004 to June 2006. We define the volatility as
the magnitude of logarithmic index return between two consecutive
minutes, that is $R(t)=|\ln Y(t)-\ln Y(t-1)|$, where the index $Y$
is the closest tick to a minute mark. Thus the sampling time is one
minute, and the volatility data size is about 140,000.

The intraday volatilities of both indices exhibit a L-shaped
intraday pattern \cite{Ni-Zhou-2007-XXX}, similar to the individual
stocks \cite{Ni-Zhou-2007-XXX,Ren-Guo-Zhou-2009-PA}. When dealing
with intraday data, this pattern should be removed
\cite{Wang-Yamasaki-Havlin-Stanley-2006-PRE,Wang-Weber-Yamasaki-Havlin-Stanley-2007-EPJB,VodenskaChitkushev-Wang-Weber-Yamasaki-Havlin-Stanley-2008-EPJB,Wang-Yamasaki-Havlin-Stanley-2008-PRE,Qiu-Guo-Chen-2008-PA}.
Otherwise, the return intervals distribution will exhibit daily
periodicity for large thresholds. The intraday pattern $A(s)$ is
defined as
\begin{equation}
  A(s)=\frac{1}{N}\sum_{i=1}^{N} R^i(s),
  \label{e10}
\end{equation}
which is the volatility at a specific moment $s$ of the trading day
averaged over all $N$ trading days and $R^i(s)$ is the volatility at
time $s$ of day $i$. The intraday pattern is removed as follows
\begin{equation}
  R'(t)=\frac{R(t)}{A(s)}. \label{e20}
\end{equation}
Then we normalize the volatility by dividing its standard deviation
\begin{equation}
  v(t)=\frac{R'(t)}{[\langle R'(t)^2 \rangle-\langle R'(t)\rangle^2]^{1/2}}.
  \label{e30}
\end{equation}

\section{Probability distribution of return intervals}
\label{S2:KS}

\subsection{Probability distribution of scaled return intervals}

We study the return intervals $\tau$ between successive volatilities
exceeding a certain threshold $q$. A series of return intervals are
obtained for each particular threshold $q$ and its number decreases
with increasing threshold $q$. For each value of $q$, we can obtain
empirically a probability distribution $P_q(\tau)$ of the volatility
return intervals, which is related to the probability distribution
$f_q(\tau/\langle\tau\rangle)$ of the scaled return intervals
$\tau/\langle\tau\rangle$ as follows
\begin{equation}
 P_q(\tau)=\frac{1}{\langle \tau \rangle} f_q ( \tau / \langle \tau \rangle ),
 \label{Eq:Pq:f}
\end{equation}
where $\langle \tau \rangle$ is the mean return interval that
depends on the threshold $q$. If the function $f_q(x)$ is
independent of $q$, there exists a universal function $f(x)$ such
that $f_q(x)=f(x)$ for different values of $q$. In other words, the
probability distributions $f_q(\tau/\langle\tau\rangle)$ of the
scaled return intervals collapse onto a single curve
$f(\tau/\langle\tau\rangle)$ and the return intervals exhibit
scaling behavior.

\begin{figure}[htb]
\centering
\includegraphics[width=8cm]{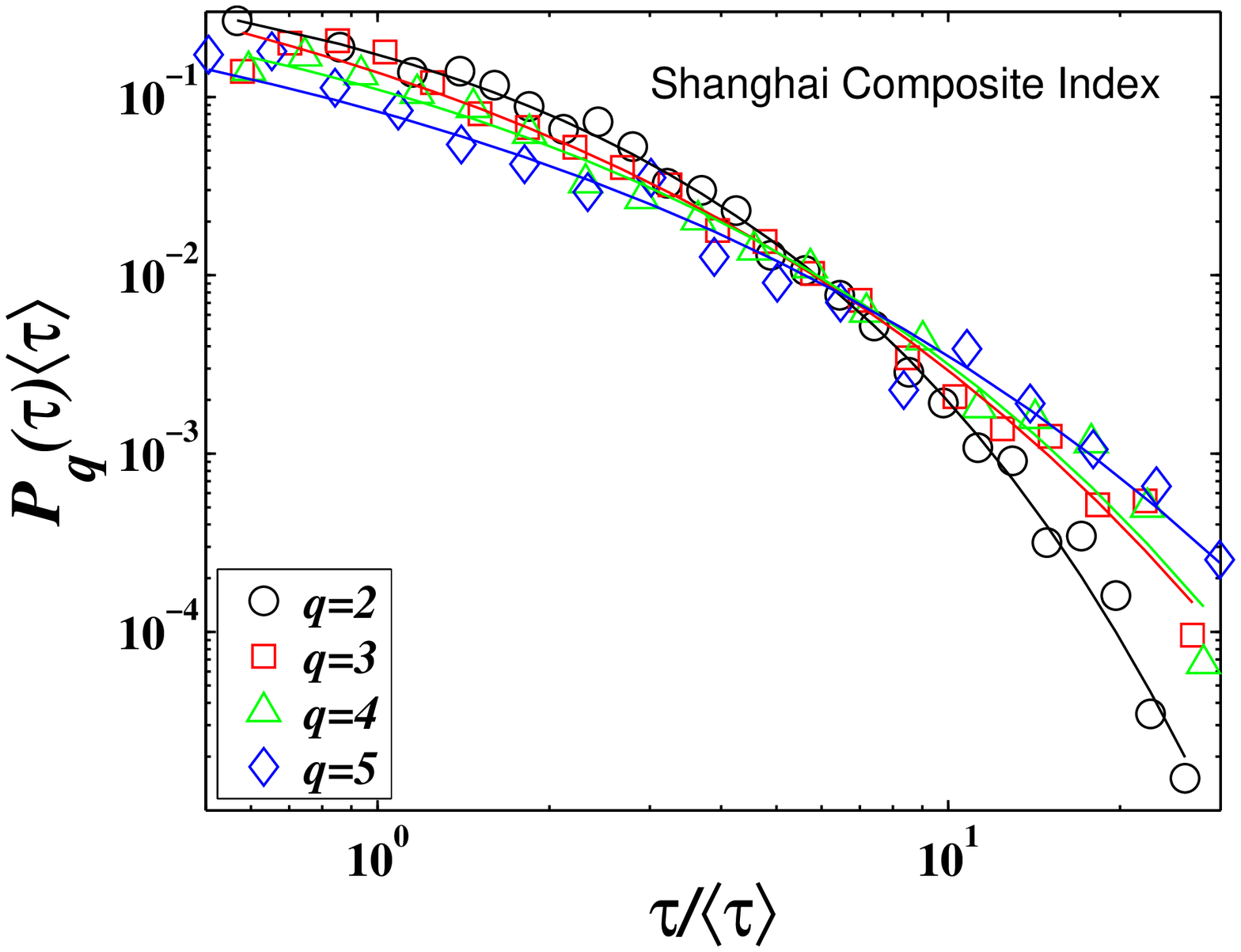}
\includegraphics[width=8cm]{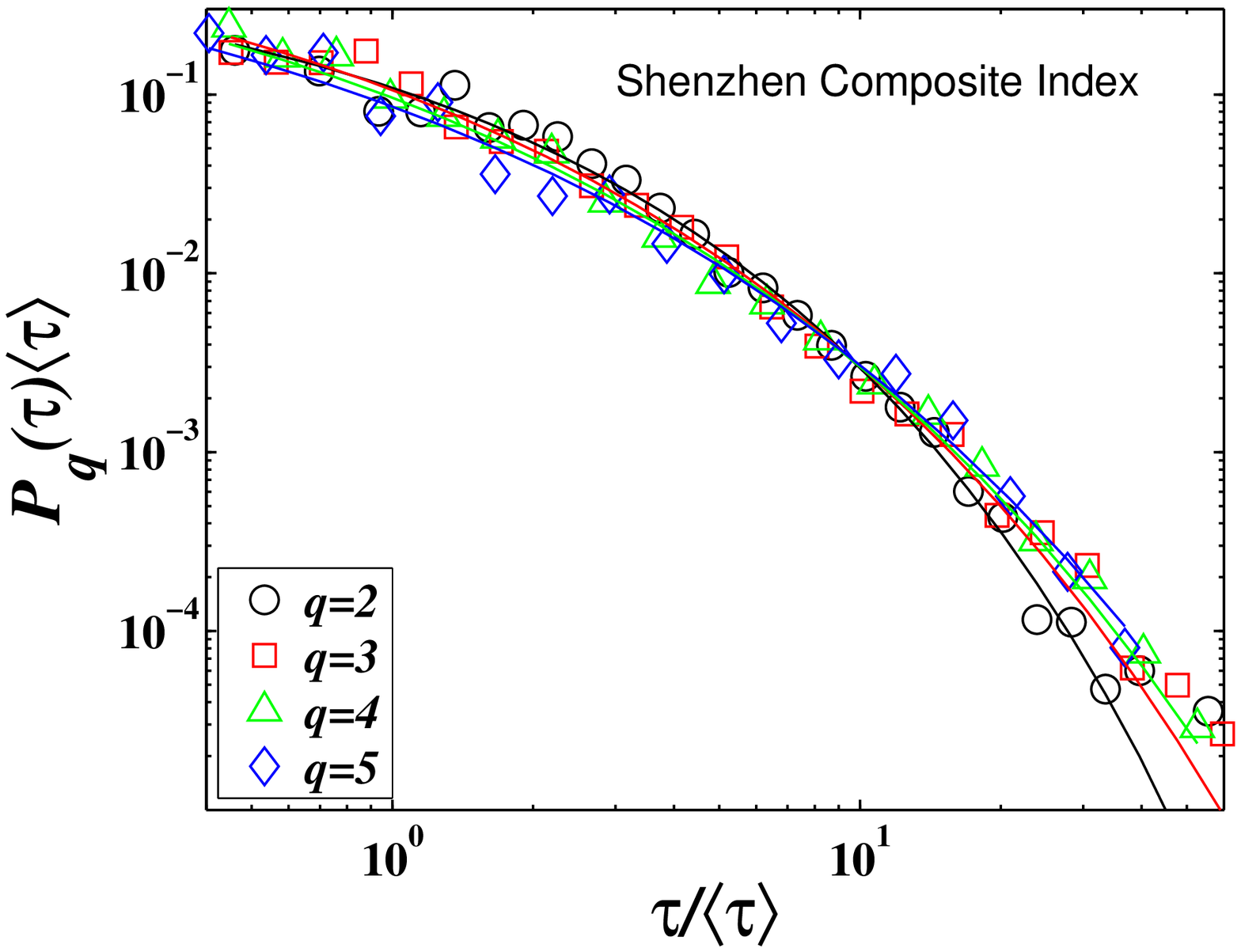}
\caption{\label{Fig:PDF:Fit} (Color online) Empirical probability
distributions of scaled return intervals for different threshold
$q=2,3,4,5$ for the Shanghai Composite Index and the Shenzhen
Composite Index. The solid curves are the fitted functions $c e^{- a
x^{\gamma}}$ with the parameters listed in
Table~\ref{TB:goodness-of-fit-KS}.}
\end{figure}

To investigate whether the return interval distributions of the two
Chinese indices exhibit scaling behavior, we plot in Figure
\ref{Fig:PDF:Fit} the empirical probability distributions
$f_q(\tau/\langle\tau\rangle)=P_q(\tau) \langle \tau \rangle$ as a
function of the scaled return intervals $\tau / \langle \tau
\rangle$ for a wide range of threshold $q=2,3,4,5$. It is evident
that the curves for different thresholds $q$ show systematic
deviations from each other and do not collapse onto a single curve,
especially for the Shanghai Composite Index. This indicates that the
distributions of return intervals for both indices could not be
approximated by a scaling relation. With the increase of the
threshold $q$, there are more large scaled return intervals and the
distribution becomes broader.

The observation that there is no scaling behavior in the volatility
return interval distributions is consistent with the results of a
previous study of individual Chinese stocks
\cite{Ren-Guo-Zhou-2009-PA}. The Kolmogorov-Smirnov test shows that
only 12 stocks out of 30 most liquid Chinese stocks exhibit scaling
behaviors in the return interval distributions for different
thresholds $q$, while the other 18 stocks do not show scaling
behavior \cite{Ren-Guo-Zhou-2009-PA}.

\subsection{Kolmogorov-Smirnov test of scaling in return interval distributions}
\label{S3:KS:Scaling}

The eyeballing of the probability distributions offers a qualitative
way of distinguishing scaling and nonscaling behaviors. Here we
further adopt a quantitative approach based on the
Kolmogorov-Smirnov test. The standard KS test is designed to test
the hypothesis that the distribution of the empirical data is equal
to a particular distribution by comparing their cumulative
distribution functions (CDFs). Our hypothesis is that the two return
interval distributions for any two different $q$ values do not
differ at least in the common region of the scaled return intervals
\cite{Jung-Wang-Havlin-Kaizoji-Moon-Stanley-2008-EPJB}. Suppose that
$F_{q_i}$ is the CDF of return intervals for $q_i$ and $F_{q_j}$ is
the CDF of return intervals for $q_j$, where $q_i\neq q_j$. We
calculate the KS statistic by comparing the two CDFs in the
overlapping region:
\begin{equation}
   KS = \max\left(|F_{q_i}-F_{q_j}|\right),~~ q_i\neq q_j~.
   \label{Eq:KS}
\end{equation}
When the KS statistic is less than a critical value $CV$, the
hypothesis is accepted and we can assume that the distribution for
$q_i$ is coincident with the distribution for $q_j$. The critical
value at the significance level of 5\% is
$CV=1.36/\sqrt{{mn}/({m+n})}$, where $m$ and $n$ are the numbers of
interval samples for $q_i$ and $q_j$, respectively.

In Table~\ref{TB:two-sample-KS} is depicted the KS statistics and
the corresponding critical values for the two indices. For the
Shanghai Composite Index, $KS>CV$ for all $(q_i,q_j)$ pairs except
$(q_i,q_j)=(4,5)$. It means that the distribution for $q=5$
coincides with the distribution for $q=4$, but significantly differs
from the distributions for other $q$ values. Similar phenomenon is
observed for the Shenzhen Composite Index. Therefore, we can
conclude that the distributions differs for different $q$ and do not
collapse onto a single curve. The KS test confirms the result that
the return interval distributions do not exhibit scaling behavior.

\begin{table}[htb]
 \centering
\caption{\label{TB:two-sample-KS} The Kolmogorov-Smirnov test of
return interval distributions by comparing the statistic $KS$ with
the critical value $CV$.}
\medskip
\begin{tabular}{llll|llll}
  \hline\hline
  \multicolumn{4}{c}{Shanghai Composite Index} & \multicolumn{4}{c}{Shenzhen Composite Index}\\  %
  \cline{1-4} \cline{5-8}
    $q_i$ & $q_j$ & $KS$ & $CV$ & $q_i$ & $q_j$ & $KS$ & $CV$\\
  \hline
    $2$ & $3$ & $0.0363$ & $0.0201$ & $2$ & $3$ & $0.0321$ & $0.0211$\\
    $2$ & $4$ & $0.0720$ & $0.0296$ & $2$ & $4$ & $0.0401$ & $0.0294$\\
    $2$ & $5$ & $0.1170$ & $0.0434$ & $2$ & $5$ & $0.0535$ & $0.0408$\\
    $3$ & $4$ & $0.0436$ & $0.0329$ & $3$ & $4$ & $0.0244$ & $0.0326$\\
    $3$ & $5$ & $0.0870$ & $0.0456$ & $3$ & $5$ & $0.0444$ & $0.0432$\\
    $4$ & $5$ & $0.0445$ & $0.0506$ & $4$ & $5$ & $0.0192$ & $0.0478$\\
  \hline\hline
\end{tabular}
\end{table}

\subsection{Fitting the return interval distributions}
\label{S4:KS:multicaling}

For those stock markets showing scaling behavior in the volatility
return interval distributions, it is a consensus that the scaling
form could be approximated by a stretched exponential function
\cite{Wang-Yamasaki-Havlin-Stanley-2006-PRE,Wang-Weber-Yamasaki-Havlin-Stanley-2007-EPJB,Jung-Wang-Havlin-Kaizoji-Moon-Stanley-2008-EPJB,Qiu-Guo-Chen-2008-PA,Wang-Yamasaki-Havlin-Stanley-2008-PRE,Wang-Yamasaki-Havlin-Stanley-2008-XXX}.
\begin{equation}
  f_q(x)=f(x)=c e^{- a x^{\gamma}},
  \label{Eq:StrExp}
\end{equation}
where $c$ and $a$ are two parameters and $\gamma$ is the correlation
exponent characterizing the long-term memory of volatilities. There
is nevertheless exceptions. Based on the KS test and the weighted KS
test, Ren, Guo and Zhou showed that the scaled return interval
distributions of 6 stocks (out of the 12 stocks exhibiting scaling
behavior) can be nicely fitted by a stretched exponential function
with $\gamma\approx0.31$ at the significance level of 5\%
\cite{Ren-Guo-Zhou-2009-PA}.

In this work, we have demonstrated that the return interval
distributions of the two Chinese indices do not follow a scaling
form. It is still interesting to check if the (scaled) return
intervals follow a stretched exponential distribution expressed in
Eq.~(\ref{Eq:StrExp}) but with different values of parameters $c$,
$a$ and the correlation exponent $\gamma$ for different threshold
$q$. In this case, our hypothesis is that the empirical distribution
is coincident with its best fitted stretched exponential function.
Similar to the KS test we have conducted for two empirical samples,
we use the KS statistics to test whether the distribution for a
certain threshold $q$ is identical to its best fitted distribution
in the overlapping region of the scaled return intervals. Let $F_q$
be the cumulative distribution for $q$ and $F_{\rm{SE}}$ the
cumulative distribution from integrating the fitted stretched
exponential. The KS statistic defined in Eq.~(\ref{Eq:KS}) becomes
\begin{equation}
   KS = \max\left(|F_q-F_{\rm{SE}}|\right),~~~~q\in \{2,3,4,5\}~.
   \label{Eq:KS2}
\end{equation}
Then the bootstrapping approach is adopted
\cite{Clauset-Shalizi-Newman-2007-XXX,Gonzalez-Hidalgo-Barabasi-2008-Nature}.
To do this, we first generate 1000 synthetic samples from the best
fitted distribution and then reconstruct the cumulative distribution
$F_{\rm{sim}}$ of each simulated sample and its CDF
$F_{\rm{sim,SE}}$ from integrating the fitted stretched exponential.
We calculate the values of $KS$ between the fitted CDF and the
simulated CDF using
\begin{equation}
   KS_{\rm{sim}} = \max\left(|F_{\rm{sim}}-F_{\rm{sim,SE}}|\right).
   \label{Eq:KS2:sim}
\end{equation}
The $p$-value is determined by the frequency that
$KS_{\rm{sim}}>KS$. The tests are carried out for the two Chinese
indices. The parameters of the fitted stretched exponential and
resultant $p$-values for different $q$ are depicted in Table
\ref{TB:goodness-of-fit-KS}.

\begin{table}[htb]
 \centering
\caption{\label{TB:goodness-of-fit-KS} The Kolmogorov-Smirnov test
of return interval distributions by comparing empirical data with
the best fitted distribution and synthetic data with the best fitted
distribution.}
\medskip
\begin{tabular}{cccccccccc}
  \hline\hline
 \multirow{3}*[2mm]{$q$}&\multicolumn{4}{c}{Shanghai Composite Index}&& \multicolumn{4}{c}{Shenzhen Composite Index}\\  %
    \cline{2-5} \cline{7-10}
      & $c$ & $a$ & $\gamma$ & $p$  && $c$ & $a$ & $\gamma$ & $p$\\
  \hline
    $2$ & $0.80$ & $ 2.05$ & $0.59$ & $0.80$ && $0.67$ & $ 3.51$ & $0.47$ & $0.02$\\
    $3$ & $2.13$ & $14.20$ & $0.38$ & $0.63$ && $1.55$ & $14.85$ & $0.37$ & $0.54$\\
    $4$ & $0.92$ & $ 5.79$ & $0.43$ & $0.58$ && $1.71$ & $22.01$ & $0.34$ & $0.72$\\
    $5$ & $1.05$ & $14.19$ & $0.35$ & $0.74$ && $1.56$ & $25.06$ & $0.33$ & $0.85$\\
  \hline\hline
\end{tabular}
\end{table}

The $p$ value could be regarded as the probability that the
empirical distribution consists with its best fit. Consider the
significance level of 1\%. If the $p$-value of an index for a
certain threshold $q$ is less than 1\%, then the null hypothesis
that the empirical PDF of this index can be well fitted by a
stretched exponential is rejected. According to
Table~\ref{TB:goodness-of-fit-KS}, the null hypotheses for all the
$q$ values are accepted for both two Chinese indices. It is
noteworthy to point out that the $p$-values for all the $q$ values
(except for $q=2$ for the Shenzhen Composite Index) are very large,
implying high goodness-of-fit of the stretched exponential to the
empirical PDFs. At the significance level of 5\%, the stretched
exponential is rejected when $q=2$ for the Shenzhen Composite Index.
To show how good the stretched exponential fits the data, we
illustrate in Figure~\ref{Fig:PDF:Fit} the fitted stretched
exponential with the parameters listed in
Table~\ref{TB:goodness-of-fit-KS}. It is obvious that the empirical
PDFs could be well fitted by a stretched exponential. In principle,
the stretched exponential fits the empirical PDF better when the
$p$-value is larger. For instance, the stretched exponential fits
the empirical PDF for the Shanghai Composite Index better than the
Shenzhen Composite Index when $q=2$.

According to Table \ref{TB:goodness-of-fit-KS}, the parameters
differ from one another, providing further evidence supporting our
conclusion that the return interval distributions do not have a
scaling form. On average, the exponent $\gamma$ decreases with
increasing threshold $q$, which is in line with the US stocks
\cite{Wang-Yamasaki-Havlin-Stanley-2008-XXX}.

\section{Moments of scaled return intervals}
\label{S5:Moment}

The distributions of return intervals exhibit multiscaling behavior
and show a systematic tendency with the threshold $q$. To further
study this tendency of the interval distribution with $q$, we
compute the moments of the scaled return intervals $x=\tau/\langle
\tau \rangle$ defined as
\begin{equation}
 \mu_m =\langle (\tau/ \langle \tau \rangle)^m \rangle ^{1/m}=\left[ \int_0^\infty x^m f_q(x) dx \right]^{1/m}.
 \label{Eq:Mom:PDF}
\end{equation}
where the mean interval $\langle \tau \rangle$ is dependent of the
threshold $q$. When $m=1$, we have $\mu_1=1$ by definition,
independent of $q$. If there is a scaling behavior that
$f_q(x)=f(x)$, the $m$-th moment $\mu_m$ is a univariate function of
the order $m$ and is independent of any other variables including
the threshold $q$ and the mean return interval $\langle \tau
\rangle$. On the contrary, the $m$-th moment $\mu_m$ is not constant
with respect to $\langle \tau \rangle$ for $m\neq1$, when there is
no scaling in the return interval distributions.

\subsection{Dependence of moment on mean return interval}
\label{S6:Moment:interval}

\begin{figure}[htb]
\centering
\includegraphics[width=8cm]{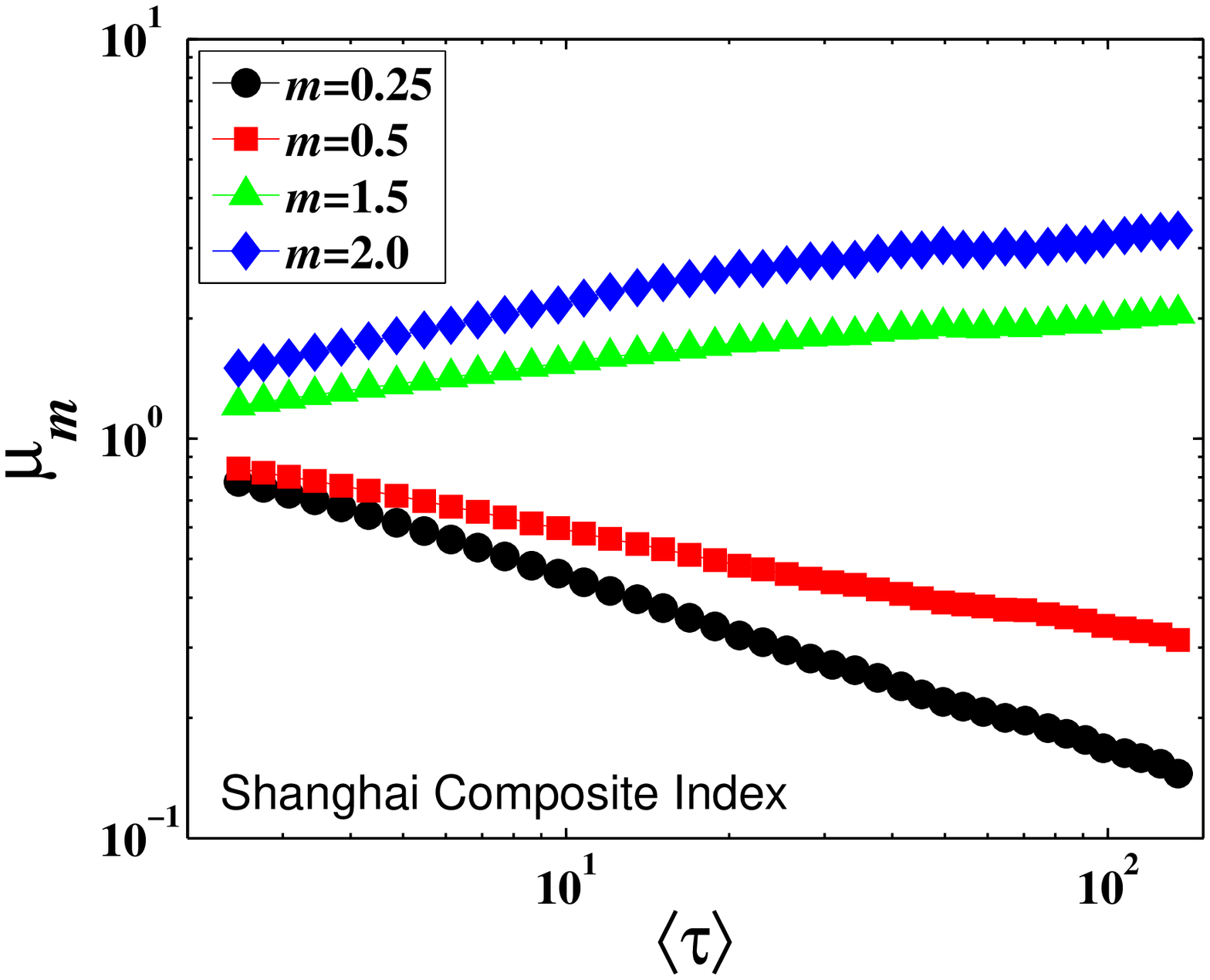}
\includegraphics[width=8cm]{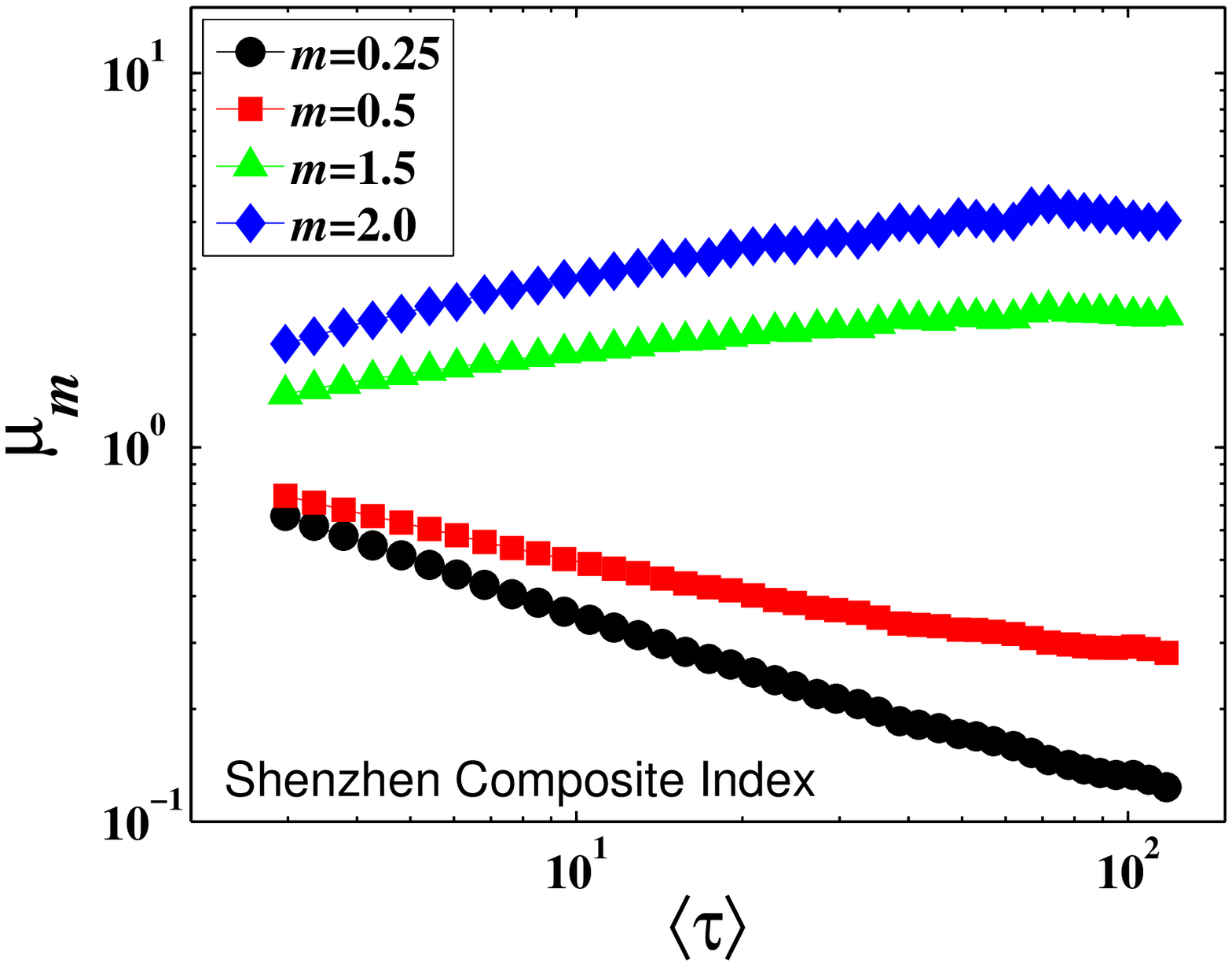}
\caption{(Color online) Moment $\mu_m$ vs $\langle \tau \rangle$ for
the two Chinese indices: Shanghai Composite Index and Shenzhen
Composite Index. For each data set, four moments $m=0.25, 0.5, 1.5,
2.0$ are presented.} \label{Fig:mom:tau}
\end{figure}

We first investigate the relation between $\mu_m$ and $\langle \tau
\rangle$. To better quantify the dependence of $\mu_m$ on $\langle
\tau \rangle$, we calculate the moments in a certain medium range of
$\langle \tau \rangle$ to avoid the finite size effect and
discreteness effect \cite{Wang-Yamasaki-Havlin-Stanley-2008-PRE}.
Figure~\ref{Fig:mom:tau} illustrates the moments $\mu_m$ for
$m=0.25, 0.5, 1.5, 2.0$ versus $\langle \tau \rangle$ for the two
Chinese indices. We investigate $\mu_m$ for a range of $\langle \tau
\rangle$ corresponding to $1 \leq q \leq 5$. Each curve of the
moments $\mu_m$ significantly deviate from a horizontal line,
confirming the multiscaling behavior in the return interval
distributions. The moment function $\mu_m$ decreases with the
increase of $\langle \tau \rangle$ when $m<1$, and shows an
increasing tendency with the increase of $\langle \tau \rangle$ when
$m>1$. These two types of moment functions are delimited by the
horizontal line $\mu_1=1$.

Take a careful look at $\mu_m$ in Figure~\ref{Fig:mom:tau}, for
$m<1$ ($m>1$) $\mu_m$ first decreases (increases) rapidly with the
increase of $\langle \tau \rangle$, and then starts to decrease
(increase) relatively slowly at $\langle \tau \rangle=10$. The
discreteness of the records of $\tau$ responds to the rapid increase
(decrease) of $\mu_m$ for small $\langle \tau \rangle$ ($\langle
\tau \rangle<10$). The moment for extremely large $\langle \tau
\rangle$, i.e., $\langle \tau \rangle>100$ ($\langle \tau \rangle$
is in units of standard deviations), will increase (decrease) for
$m<1$ ($m>1$) due to the finite size effect. We choose to study
$\mu_m$ in a medium region $10<\langle \tau \rangle<100$ where in
the effects of finite size and discreteness are small and $\mu_m$
shows a clear power-law-like trend with $\langle \tau \rangle$. We
use a power law to fit the moment in this medium region,
\begin{equation}
   \mu_m \sim \langle \tau \rangle^{\alpha}.
   \label{Eq:moment:PL}
\end{equation}
If the PDF of return intervals follow a scaling form, $\mu_m$ is
independent of $\langle \tau \rangle$ according to
Eq.~(\ref{Eq:Mom:PDF}) and the exponent $\alpha$ should be some
value very close to $0$. If the exponent $\alpha$ is significantly
different from $0$, it implies that the PDF of return intervals may
show multiscaling behavior.

\begin{figure}[htb]
\centering
\includegraphics[width=8cm]{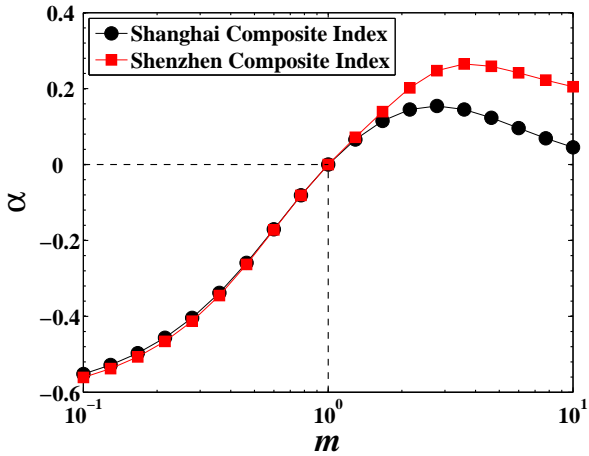}
\caption{(Color online) Exponent $\alpha$ of $\mu_m$ in the region
$10<\langle\tau\rangle<100$ for the two Chinese indices: Shanghai
Composite Index and Shenzhen Composite Index presented by circles
and squares respectively.} \label{Fig:mom:slope}
\end{figure}

Figure~\ref{Fig:mom:slope} plots the exponent $\alpha$ as a function
of order $m$ for the two Chinese indices. This figure shows that the
exponent $\alpha$ differs from 0 in a systematic fashion. The
exponents for the two indices are very close to each other when $m$
is small. For $m<1$, $\alpha$ is negative. The exponent $\alpha$
increases with $m$ when $m<3$ and decreases afterwards owning to the
finite size effect. For large order $m$, the $\alpha$ value for the
Shenzhen Composite Index is greater than that for the Shanghai
Composite Index. This implies that large $\langle \tau \rangle$
tends to occur with greater probability for the Shenzhen Composite
Index than the Shanghai Composite Index, since large $\langle \tau
\rangle$ contributes more for high order $\mu_m$.

The above analysis according to Eq.~(\ref{Eq:moment:PL}) can be
related to the extended self-similarity (ESS) analysis
\cite{Benzi-Ciliberto-Tripiccione-Baudet-Massaioli-Succi-1993-PRE},
which reads
\begin{equation}
   \langle \tau^m \rangle \sim \langle \tau^n \rangle^{\xi(m,n)}.
   \label{Eq:ESS}
\end{equation}
If the generalized variable of $\mu_m$
\begin{equation}
 \mu_{m,n} =\left\langle \left(\frac{\tau}{ \langle \tau^n \rangle^{1/n}}\right)^m \right\rangle ^{1/m}
  =\frac{\langle \tau^m \rangle^{1/m}}{ \langle \tau^n \rangle^{1/n}}
 \label{Eq:mu:mn}
\end{equation}
scales as
\begin{equation}
 \mu_{m,n} \sim  \left(\langle \tau^n \rangle^{1/n} \right)^{\alpha}
 \label{Eq:mu:mn:alpha}
\end{equation}
together with Eq.~(\ref{Eq:ESS}), we have
\begin{equation}
   (\alpha+1)/n = \xi(m,n)/m.
   \label{Eq:ESS:exponents}
\end{equation}

If the return interval distribution can be scaled as follows
\begin{equation}
 P_q(\tau)=\frac{1}{\langle \tau^n \rangle^{1/n}} f \left( \frac{\tau}{ \langle \tau^n \rangle^{1/n}} \right),
 \label{Eq:P:f}
\end{equation}
we obtain that
\begin{equation}
   \xi(m,n) = m/n.
   \label{Eq:ESS:xi:mn}
\end{equation}
In this case, we have
\begin{equation}
   \alpha = 0.
   \label{Eq:ESS:alpha:0}
\end{equation}
In other words, $\mu_{m,n}$ is independent of $\langle \tau^n
\rangle^{1/n}$.

Our empirical test focuses on the case that $n=1$. This ESS
framework was also used to investigate scaling in the exit times in
turbulence \cite{Zhou-Sornette-Yuan-2006-PD} and intertrade
durations \cite{Eisler-Kertesz-2006-EPJB}. According to
Eq.~(\ref{Eq:ESS:exponents})
\begin{equation}
   \alpha(m) = \xi(m,1)/m-1.
   \label{Eq:ESS:alpha:xi:m1}
\end{equation}
Since $\xi(1,1)=1$, we have $\alpha(1)=0$ when $m=1$. This is well
verified by Figure \ref{Fig:mom:slope}.

\subsection{Dependence of moment on order m}
\label{S7:Moment:order}

\begin{figure}[b!]
\centering
\includegraphics[width=7cm]{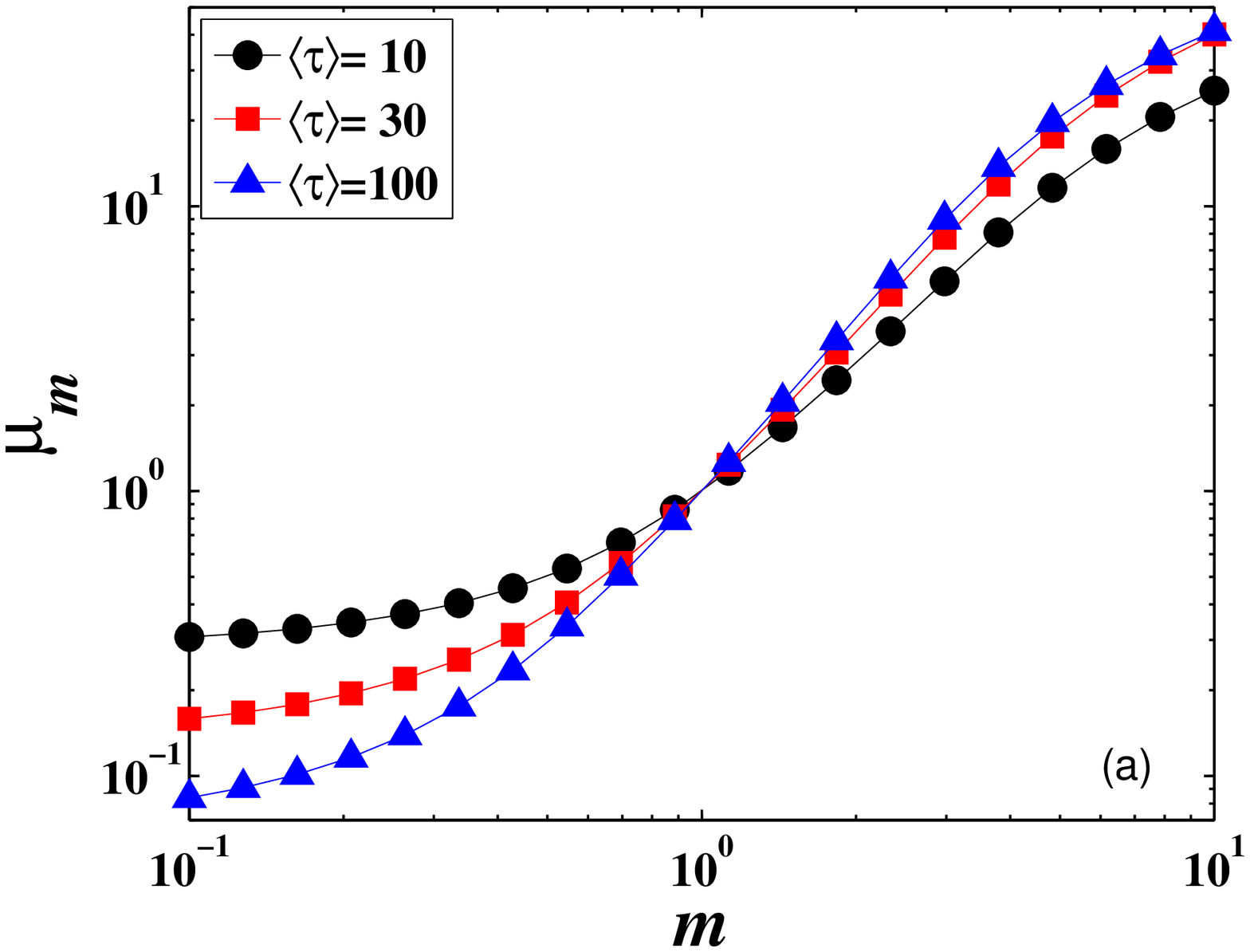}
\includegraphics[width=7cm]{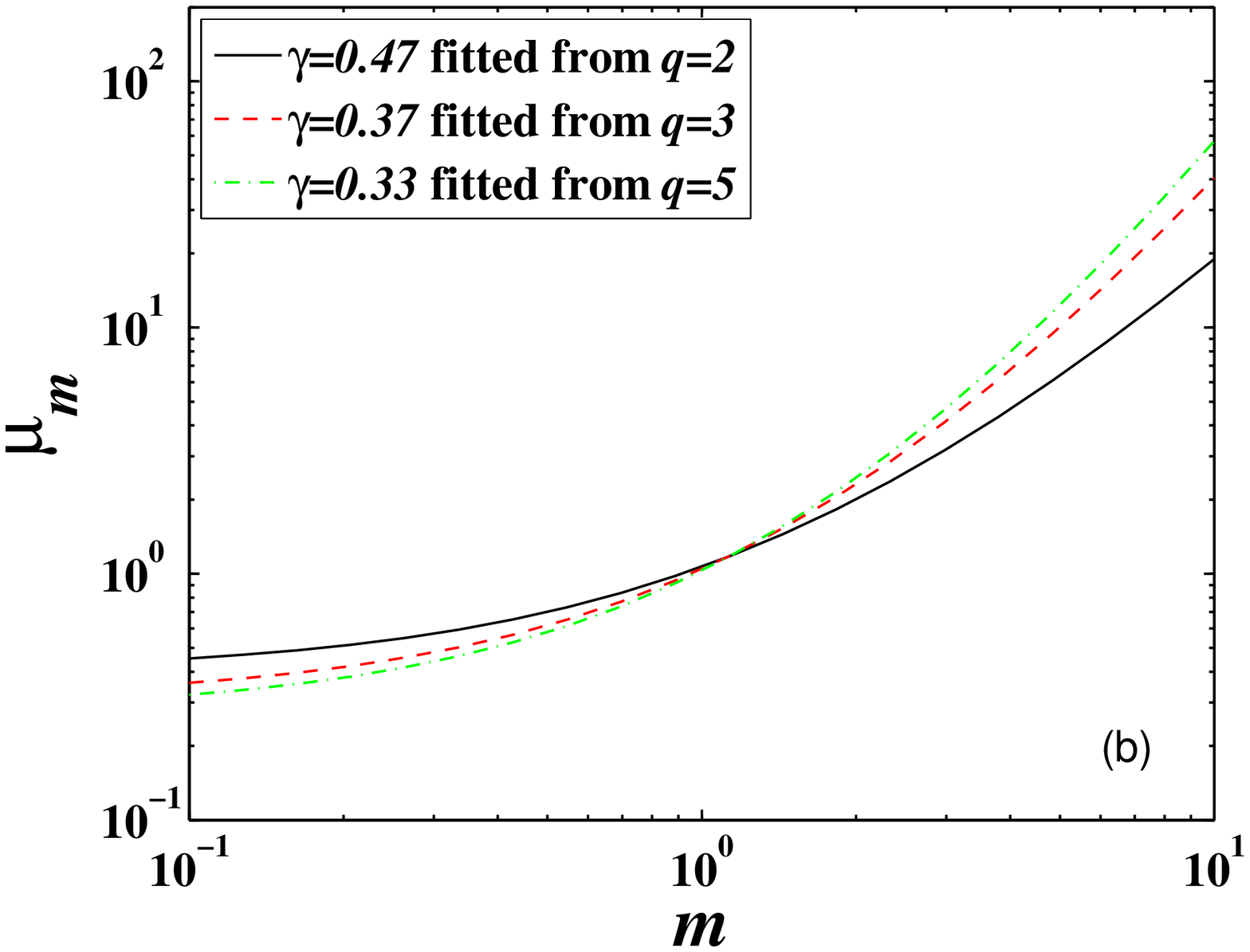}
\caption{(Color online) (a) Moment $\mu_m$ vs $m$ for Shenzhen
Composite Index for $\langle\tau\rangle=10,30,100$ corresponding to
$q=2.0,3.2,4.8$, presented by circles, squares, triangles
respectively. (b) Analytical moments obtained from the stretched
exponential distributions with parameters fitted from empirical data
of Shenzhen Composite Index for $q=2,3,5$.} \label{Fig:mom:order}
\end{figure}

The moment $\mu_m$ not only display a significant dependence on
$\langle \tau \rangle$, but also shows a systematic tendency with
$m$ as shown in Figure~\ref{Fig:mom:tau}. It is interesting to
investigate the relation between $\mu_m$ and $m$ directly. For a
fixed $\langle \tau \rangle$, one can study the moment of $\tau$ of
various orders $m$. If the return interval distribution strictly
obeys a scaling form, $\mu_m$ should not depend on $\langle \tau
\rangle$, and the $\mu_m$ curves for different $\langle \tau
\rangle$ should all collapse onto a single curve. We plot in
Figure~\ref{Fig:mom:order}(a) the moment $\mu_m$ as a function of
$m$ for fixed $\langle \tau \rangle=10,30,100$ corresponding to
$q=2.0,3.2,4.8$ for the Shenzhen Composite Index. One sees the
curves for different $\langle \tau \rangle$ exhibit substantial
deviations from a single curve, which demonstrates the multiscaling
behavior of return intervals. For small $m$ ($m<1$), $\mu_m$
decreases with the increase of $\langle \tau \rangle$, while for big
$m$ ($m>1$) $\mu_m$ tends to increase with the increase of $\langle
\tau \rangle$. This is not difficult to understand since small
$\tau$ dominates $\mu_m$ for small order $m$ and large $\tau$
dominates $\mu_m$ for large order $m$. The situation for the
Shanghai Composite Index is very similar.

We have demonstrated that the return interval distribution follows a
stretched exponential form with different parameters for various
thresholds $q$ in Section~\ref{S4:KS:multicaling}. For the data
perfectly follow a stretched exponential distribution, we can
calculate the analytical result of the moment $\mu_m$ by
substituting Eq.~(\ref{Eq:StrExp}) to Eq.~(\ref{Eq:Mom:PDF}) and
considering the normalization condition of probability density. It
follows immediately that
\cite{Wang-Yamasaki-Havlin-Stanley-2008-PRE}
\begin{equation}
   \mu = \frac{1}{a} \left[ \frac{\Gamma((m+1)/\gamma)}{\Gamma(1/\gamma)} \right]^{1/m} ~.
   \label{Eq:Mom:Ana1}
\end{equation}
In Figure~\ref{Fig:mom:order}(b), the analytical curves of $\mu_m$
versus $m$ for three stretched exponential distributions fitted from
empirical data of Shenzhen Composite Index for $q=2,3,5$ are
plotted. As one can see that the analytical results are similar to
that of the empirical data, which supports the multiscaling of the
empirical return intervals.

\section{Summary and conclusions}
\label{S8:concl}

We have studied the multiscaling properties of the distributions of
volatility return intervals for two Chinese indices, together with
their moments. The Kolmogorov-Smirnov test is adopted to examine the
scaling behavior of the return interval distributions as well as the
particular form of the distribution. We find that the return
intervals of the two indices exhibit multiscaling behaviors, and
their distributions for different thresholds $q$ can be well
approximated by stretched exponential functions $f_q(x) \sim e^{- a
x^{\gamma}}$ but with different values of the correlation exponent
$\gamma$. An ESS-like moment analysis confirms the existence of
multiscaling rather than monoscaling. This result is consistent with
previous analysis on individual Chinese stocks
\cite{Ren-Guo-Zhou-2009-PA} and help us better understand the
properties of volatility return intervals in the Chinese stock
markets.

\begin{acknowledgement}
This work was partially supported by the Shanghai Educational
Development Foundation (No. 2008CG37), the National Natural Science
Foundation of China (No. 70501011), the Fok Ying Tong Education
Foundation (No. 101086), and the Program for New Century Excellent
Talents in University (No. NCET-07-0288).
\end{acknowledgement}

\bibliographystyle{epj}
\bibliography{E:/paper/bibfile/Bibliography}

\end{document}